\documentclass{sigchi}

\CopyrightYear{2016}
\setcopyright{acmlicensed}
\doi{http://dx.doi.org/10.475/123_4}
\isbn{123-4567-24-567/08/06}
\conferenceinfo{CHI'16,}{May 07--12, 2016, San Jose, CA, USA}
\acmPrice{\$15.00}

 \toappear{
	Submitted to UIST 2020.
 }

\usepackage{balance}       
\usepackage{graphics}      
\usepackage[T1]{fontenc}   
\usepackage{txfonts}
\usepackage{mathptmx}
\usepackage[pdflang={en-US},pdftex]{hyperref}
\usepackage{color}
\usepackage{booktabs}
\usepackage{textcomp}
\usepackage{xac}
\usepackage{mathtools}
\usepackage{float}
\usepackage{array}
\usepackage{soul}

\usepackage{microtype}        
\usepackage{ccicons}          

\usepackage{todonotes}

\def\plaintitle{Romeo: A Design Tool for Embedding Transformable Parts in 3D Models to Robotically Augment Default Functionality
}

\def\emptyauthor{}
\def\plainkeywords{Design tool; transformables; robotic task; generative design.}

\makeatletter
\def\url@leostyle{%
  \@ifundefined{selectfont}{
    \def\UrlFont{\sf}
  }{
    \def\UrlFont{\small\bf\ttfamily}
  }}
\makeatother
\def\pprw{8.5in}
\def\pprh{11in}

\definecolor{linkColor}{RGB}{6,125,233}
\hypersetup{%
  pdftitle={\plaintitle},
  pdfauthor={\emptyauthor},
  pdfkeywords={\plainkeywords},
  pdfdisplaydoctitle=true, 
  bookmarksnumbered,
  pdfstartview={FitH},
  colorlinks,
  citecolor=black,
  filecolor=black,
  linkcolor=black,
  urlcolor=linkColor,
  breaklinks=true,
  hypertexnames=false
}

\begin{document}

\title{\plaintitle}

\numberofauthors{4}
\author{%
 \alignauthor{Jiahao Li\\
    \affaddr{UCLA HCI Research}\\
    \email{ljhnick@g.ucla.edu}}\\
 \alignauthor{Meilin Cui\\
    \affaddr{UCLA HCI Research}\\
    \email{ecui17@g.ucla.edu}}\\
 \alignauthor{Jeeeun Kim\\
    \affaddr{CS\&E Texas A\&M University}\\
    \email{jeeeun.kim@tamu.edu}}\\
 \alignauthor{Xiang `Anthony' Chen\\
    \affaddr{UCLA HCI Research}\\
    \email{xac@ucla.edu}}\\
}
\teaser{
   \centering
   \includegraphics[width=2.1\columnwidth]{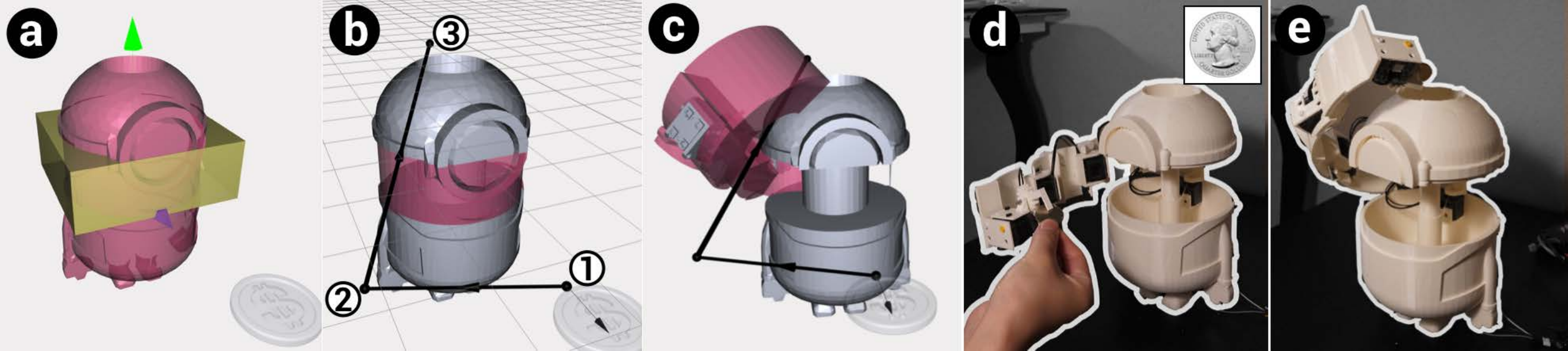}
   \caption{
   Romeo enables a user to transform a Minion model into a coin-stealing piggybank by selecting the mid-section of the Minion as the transformable part (a), specifying motion path for the transformed part to pick up a coin and place it in the bank (b), generating the corresponding transformable robotic arm (c) and 3D printing and installing the result (d-e).
   }~\label{fg:fig1}
   \vskip -8pt
}

\maketitle

\begin{abstract}
  Reconfiguring shapes of objects enables transforming existing passive objects with robotic functionalities,
\eg a transformable coffee cup holder can be attached to a chair's armrest, a piggybank can reach out an arm to 'steal' coins. 
Despite the advance in end-user 3D design and fabrication, 
it remains challenging for non-experts to create such `transformables' using existing tools due to the requirement of specific engineering knowledge such as mechanisms and robotic design. 

We present \textit{Romeo}---a design tool for creating transformables embedded into a 3D model to robotically augment the object's default functionalities. 
Romeo allows users to express
at a high level, (1) which part of the object to be transformed, (2) how it moves following motion points in space, 
and (3) the corresponding action to be taken.
Romeo then automatically generates a robotic arm embedded in the transformable part ready for fabrication. 
We validated Romeo with a design session where 8 participants design and
create custom transformables using
3D objects of their own choice. 
\end{abstract}

\ccsdesc[500]{Human-centered computing~Interactive systems and tools}

\keywords{\plainkeywords}
\printccsdesc


\section{Introduction}


Objects that can transform its geometry and/or functionality (which we referred to as transformables)
hold the promises of dynamically adapting to multiple usages by reconfiguring shapes, either automatically or via manual reconfiguration as depicted in \fgref{existing} with several examples.


The advent of computational design and 3D modelling tools offers the possibilities for casual makers \cite{hudson2016understanding} to create custom objects. 
However, for non-technical users, designing transformables remains challenging due to the requirements of expert-level engineering knowledge such as mechanisms and robotic design that are not provided in existing end-user oriented design tools. 

\fgh{existing}{existing_examples}{1}{Existing examples of objects with multiple functionalities by transforming its geometry: a stealing coin piggybank (a), a transformable cup holder (b), and an iPad cover folded into a stand (c).
}



Prior work focused on \emph{actuating} passive objects \cite{chen2016reprise,ramakers2016retrofab,DBLP:conf/uist/LiKC19} using attachable mechanisms, yet has not taken account how to \emph{transform} given objects. 
Prior work that addressed transformables tends to focus on computational analysis of geometry and optimization \cite{yuan2018computational, Boxelization, HUANG2016127}, rather than providing design assistance to create user-defined custom transformables. 
Further, these approaches almost exclusively exploit geometry as the only constraint when generating a transformable design (\eg \cite{yuan2018computational}), with little consideration of obtained functionality, \ie how a transformable can perform a user-defined task.

We present Romeo---a design tool for generating and embedding transformable parts in 3D models to robotically augment the default functionality. \fgref{fig1} shows an exemplar workflow of Romeo: A user selects a cross section of the minion to be a transformable part (a), specifies a motion path to pick up a coin and place it in the bank (b), based on which Romeo generates the corresponding transformable robotic arm (c), which is then 3D printed and installed with motors to perform the coin-stealing task (d-e).



We conducted a design session with eight participants conducting both controlled task to replicate a task using given 3D objects (stirring spatula and sanitizing tissue box) and open-ended tasks (design a piggy bank using participant-chosen 3D model) of creating transformables.
Overall, participants were able to understand and complete design using Romeo as a design tool, to define functional tasks by parts selection and motion path specification, to generate ready-to-print transformable parts for fabrication, and to assemble the parts into a fully functional transformable object. 


{\bf Our main contribution} is an end-user design tool that addresses {\it both} geometric and functional constraints in generating a transformable part embedded into an object's 3D model (geometric constraint) while being able to transform and perform a user-defined task (functional constraint).


\section{Related Work}
Romeo provides end-users with a design tool to augment an object's functionality by transforming a subset of the object. This goal cross-cuts three areas of prior work: \one designing objects that can transform original shapes
\two reality-based design tools that extract real-world information (e.g., object geometry and context) to create designs; and 
\three interactive design of robotic characters.

\subsection{Computational Design of Shape-Changing Objects}
Past research has explored the design of articulated objects that can actuate existing shape by transformation.
Cal{\`\i} \etal \ proposed to generate assembly-free models by inserted joint configurations of an animation rig to articulate mechanical object \cite{cali20123d}. B{\"a}cher \etal \ takes a skinned mesh as input and estimate the corresponding virtual articulated part segment to compute placement of the joints \cite{bacher2012fabricating}. Ureta \etal \ proposes an interactive system for creating 3D printable joints with user-controlled appearance while taking into account the range of motion achievable \cite{ureta2016interactive}. All this work focuses largely on inserting assembly-free joints in the articulated parts given the input shapes, rather than to transform such objects away from their given shapes. 

One body of research on transforming shapes is concerned with
the design of reconfigurable objects.
Li \etal \ and Garg \etal \
proposed computational approaches to design foldable furniture to save space while not in use \cite{li2015foldabilizing, garg2016computational}. 
In this process, Boxelization helps transform a 3D object into a series of small cubes where adjacent cubes are either linearly-linked and/or fold \cite{Boxelization}. 
Huang \etal \ explores the design and animation of the motion of transformation based on the input 3D model and target skeleton representing the desired figure
\cite{HUANG2016127}. 
Yu \etal \ investigates transforming the object with telescopic structures using user sketches or an arbitrary mesh as input \cite{yu2017computational}. 
Perhaps the most related to our work is the design approach by Yuan \etal, which uses the target model and skeleton as user input to automatically generate fabricable transformable objects
\cite{yuan2018computational}. 
However, Yuan \etal \ only addresses the automatic generation of transformables based on input geometry constraints, but does not consider higher-level goals of the desired function as input. Across prior work, there is a lack tool support for non-expert users to create custom transformable objects. 

Also, design of origami robots is also related as such robots can be initially fabricated as flat sheet then be folded into a complex 3D shapes. Schulz \etal \ proposed an end-to-end system for design of robots with ground locomotion \cite{schulz2017interactive}. Mehta \etal \ proposed a tool that users can create printable 3D origami-inspired robot from high-level structural and functional specification \cite{mehta2016design, mehta2015integrated, mehta2018robot}. 
Romeo focuses on using existing 3D shapes with a default function and transform into another 3D shape with augmented functionality.




\subsection{Reality-based Design Tools}
Romeo enables end-users to define a task that the transformable part will conduct, to interact with real-world objects (\eg picking up and depositing a coin). Existing work has explored reality-based design tools that address \one how to design objects with kinematic and robotic features from \two high-level, real world context-based design goals as an input, while leaving low-level functional considerations to a generative algorithm.

Design of objects with kinematic feature that could physically interact with the real world has been a focus of many research. Obtaining an understanding of the physical world informs new ideas of augmentation, such as `mechanical hijacking' first demonstrated by Davidoff \etal \ facilitated physical interaction with real-world object using LEGO MindStorm toolkit \cite{davidoff2011mechanical}. 
TrussFormer enables users to design and 3D print large-scale structures with kinematic features \cite{kovacs2018trussformer}. 
Grafter helps end users extract and reconstruct mechanical elements from 3D printable machines. 
It affords users to be aware of real-world constraints when fabricated, allowing to clean out sweeping space along with rotation axles
\cite{roumen2018grafter}. 
RetroFab offers an authoring tool to scan an existing physical interface and automate its controls with an enclosure consisting of mechanical widgets and electronic devices \cite{ramakers2016retrofab}. 
Robiot turns legacy static objects into robots by generating 3D printable attachment mechanisms to automate physical tasks \cite{li2019robiot}. 
Similarly, Romeo aims at empowering end-users to create transformable parts of objects that are able to physically interact with real-world objects, given context-aware, user-defined task.

Generative design is also related to our work in that it allows users to only provide high-level information as input. 
Reprise invites users to express what type of action is applied to an object at a high level, and so generates adaptation for hand-held objects for easy manipulation \cite{chen2016reprise}. 
AutoConnect promotes the automatic generation of the attachment mechanisms between two user-selected objects based on input 3D model or scanned digital model of real-world object \cite{koyama2015autoconnect}. 
Further incorporate users’ intent, Forte loops user input into the optimization process to create structures that meet functional requirement as well as mimic users’ sketches \cite{chen2018forte}. 
Patching provides a hybrid platform that scans, mills, and fabricates new components ad-hoc, to replace part of an existing object with updated user-context \cite{teibrich2015patching}. 
DreamSketch allows a user to integrate generatively designed components with the workflow of sketching \cite{kazi2017dreamsketch}. 
In comparison, Romeo employs a hybrid approach that combines the automation of robotic feature generation as well as a generative process: users input a 3D model and specify a reality-based target task, based on which the tool generates the corresponding transformable parts. 

\subsection{Interactive Design of Robotic Characters}
The eventual goal of Romeo is to generate functional ready-to-transform objects based on user-defined high-level specification. 
To achieve this, Romeo builds upon prior work that examine interactive design of robotic characters. 


Megaro \etal \ present a design system that automates the tedious process of creating 3D-printable robotic creatures while allowing for customization for casual users \cite{megaro2015interactive}. Recent research focused on computational approaches for non-expert users to design animated or robotic characters by high-level motion specification such as motion curves \cite{coros2013computational, thomaszewski2014computational, ha2018computational}.
LinkedIt enables end-user design and fabrication of robots, from creating linkages to exhibit specific motion path \cite{bacher2015linkedit}.
Geppeto is an interactive system that allows users to design expressive robots by editing complex parameters in a semantic level \cite{desai2019geppetto}. 
Using modular electromechanical components known as `computational abstractions', novices become capable to easily create custom robotic devices using a visual design environment \cite{desai2017computational, desai2018automatic}. 
In contrast, Romeo enables the generation of robotic mechanisms by only requiring users to specify sample points along a motion path and to select an action to be taken along the way, thus simplifying the specification of tasks into just a few steps.



\fg{interface}{interface}{1}{A screenshot of Romeo's user interface: a) target object; b) reference object; c) functional buttons, from left to right: selecting transformable part, specifying motion points and action, generate embedded robotic arm, animation, export and d) button to restart the current step}

\section{Overview of Design and Fabrication Process}

We break down Romeo's process of generating a user-defined transformable part into four steps, which we briefly describe here and discuss in further details in the next few sections:

\begin{enumerate}
    \item \textbf{Selecting which part of the object to be transformed}. \ To start, Romeo allows users to select a part of an 3D object to be transformed by sweeping its cross-sectional area along one of the X/Y/Z axes. 
    \item \textbf{Specifying motion points to follow and corresponding action to be taken}. \ 
    Romeo lets users specify a task that consists of a series of motion points for 
    the end-effector of the robotic arm to follow, as well as 
    what action should be taken at each motion point: \one pick or place, \two follow a trajectory and \three attach to a surface.
    \item \textbf{Generating a robotic arm embedded in the transformable part}. \ Romeo then generates a robotic arm that follows the user-specified motion points to perform a task, segmenting user-specified parts into a series of joints. The object-embedded arm is visualized and the resulting motion is animated so users can iteratively modify their design.
    \item \textbf{Generating components for fabrication and deployment}. \ Finally, Romeo generates fabrication-ready 3D models, guides to assemble all the components along with motors, and software needed to control the motors that actuate the robotic arm.
\end{enumerate}

\subsection{Preprocessing}
Before the workflow starts, Romeo assumes the input 3D model has been preprocessed using existing CAD tools:
\one the model has been oriented to be as axis-aligned as possible, \eg a spatula's handle aligned with one of the X/Y/Z axes, a minion model at an upright orientation;
\two non-transformable functionality has been implemented in the model, \eg a model for piggybank has been hollowed for storing coins;
\three optionally, reference objects have been placed around the model with its relative position, \eg a pot placed next to a spatula to serve as a reference when a user specifies motion points of the spatula stirring in the pot.

\section{\#1 Selecting Part of An Object to Transform}

\fg{transformable}{transformable_part}{1}{Romeo enables selecting part of an object to transform by sweeping cross-sectional area along X/Y/Z axis.
}

Currently Romeo supports selecting a transformable part as a cross-sectional area swept along one of the X/Y/Z axes of the object with its bounding box (\fgref{transformable}).
Such an axis-aligned selection approach is designed to simplify the 3D manipulation task for non-expert users.
Admittedly, it trades off expressiveness, \eg selecting a semantically-meaningful part unaligned with X/Y/Z such as minion's goggle. which we further discuss later as future work. 


Once an axis is selected by clicking an arrow parallel to each axis, users could adjust the starting and ending position of the sweeping area, by dragging the two corresponding surfaces
of the bounding box highlighted in yellow (\fgref{transformable}b). 
A boolean operation is conducted to obtain the intersection of the target 3D object and the selected cuboid.
This intersection is then set to the transformable part and the rest is automatically set to the static part (\fgref{transformable}c). 


One issue in this step is that the transformable part is likely to divide the object into two disjoint components in the static part, sometimes causing overhang due to a vertical segmentation (\eg \fgref{transformable} c).
Romeo can detect such cases by comparing the transformable part's and the entire object's bounding boxes.
If disjoint components are detected, Romeo generates a cylinder in between as a connection (d).
The radius of the cylinder is proportional to the dimension of the transformable part, while constrained by the need to leave sufficient space for mounting motors around this pillar.

%


\section{\#2 Specifying Motion Points and Action}
The next step is to specify the motion points of the selected transformable part, in order for the parts to follow and operate the corresponding action. 
The transformation is led by the end-effector---a component that follows the path defined by user-specified motion points. 
For example, the tip of the transformed arm of the piggybank is the end-effector that will follow the motion points to steal the coin (\fgref{end-effector}).

Motion points involve two types of information: 3D positions and orientations (the orientation of the end-effector when it approaches a point). At each motion point, the end-effector will take the action in one of the three types: pick-and-place, following a trajectory or attaching to a surface.

\textbf{Position} \ One well-known challenge here is letting users directly specify a 3D point on a 2D screen. 
Romeo addresses this by providing a two-step
process.
First, the user specifies the 2D position of a motion point on a reference plane coplanar to the cross section of the transformable part (\fgref{visual_aid}a): for the very first motion point, the reference plane cuts across the centroid of the transformable part and for the subsequent points, the reference plane cuts across the previous motion point.
Next, the user specifies the third dimension by sliding the motion point along a reference line perpendicular to the reference plane (\fgref{visual_aid}c).
%
To help user have a better understanding about the relative position of motion points in 3D space, a top view and a side view are displayed on the right side of the tool (\fgref{visual_aid}b-d). While the end-effector tracks the motion points in sequence, users could specify a loop by setting the last point close to the first point (the threshold is 50 mm).

\fg{specify_points}{points}{1}{Users can specify the motion points for the end-effector to follow in order (abc) and the corresponding action to be taken (d).}
\fg{visual_aid}{visual}{1}{Romeo allow users to first specify the 2D position on a reference plane (ab), and then the third dimension along a perpendicular reference line (cd). A spherical widget is used for specifying the end-effector orientation (d). Lack of orientation specification may result in bad result (fg).}
\textbf{Orientation} \ As shown in \fgref{visual_aid}e-g, we use a spherical widget similar to \cite{chen2016reprise}, to help a user specify the orientation of the end-effector as it approaches one motion point from the previous motion point. 
Orientation is only required in the special context, for example, the contact face of the stamp needs to be parallel to the paper placed on the ground, while stamping.
Users can click on the spherical surface to specify the orientation, or by clicking anywhere outside to indicate that no specific orientation is needed.

At each motion point, the end-effector can perform one of the following three actions:
\begin{itemize}
    \item \textbf{Picking/placing} \ Picking an object and placing it at another location is the most common task for a robotic manipulator useful in many real-world contexts, \eg in the assembly line.
    The first choice of this action automatically becomes a \emph{picking} action and the following choice will be regarded as a \emph{placing} which is subsequent to the prior action.
    \item \textbf{Following a trajectory} \ indicates that the end-effector simply moves to motion point without taking any other action. For examples, additional motion points between pick and place are simply trajectory following.
    \item \textbf{Attaching to a surface} \ can be used to transform an object to attach some parts to an existing physical object, \eg coffee cup holder (\fgref{attach}a) that can be mounted on a chair's armrest. Romeo provides three standard types of surface to represent the existing geometry: cylinder, rectangular-prism and flat-plane \cite{koyama2015autoconnect} to specify an attachment surface. 
    A user places a surface object similar to specifying the position of a motion point, then the user can further specify the orientation of the surface using a spherical widget (\fgref{attach}b). 
\end{itemize}

\fg{attach}{attach}{1}{Romeo lets a users to specify the target surface for attaching action (a) and further define the orientation of the surface (b)}


\section{\#3 Generating An Embedded Robotic Arm}
As the user finished defining the task, Romeo takes the transformable part and motion points as parameters to generate an embedded robotic arm. 
The union of all the points reachable by a robotic arm is found and called as workspace, determined by linkage length and the moving range of each joint. 

One challenge here is the trade-off between embedding more links to enlarge the workspace and securing space to host these links the limited volume of selected part. 
We found an optimal balance for Romeo to segment the part into a four-joint robotic arm, 
as with four joints, Romeo could change the moving direction of each joint to cover a wide variety of user-defined tasks. 
We determined the number of joints to be four after investigating several pilot examples.
We define different combinations of each joint as configurations of the robotic arm.


Another consideration for a user to decide is at generating a robotic arm, to determine the \emph{base} and the \emph{end-effector}.
In the case of the coin-stealing piggybank (\fgref{end-effector}b), it is natural to use the static part as the base and the end of the transformable part as the end-effector for pick/placing coins.
However, in some cases the relationship needs to be flipped. For example, for a spatula (\fgref{end-effector}a), the static part (\ie the blade) becomes the end-effector as it carries the stirring function requisite for the user-specified task.
In Romeo, a user can click a button (third from left in \fgref{interface}c) to switch between two types of transformation, using the static part as the base {\it vs}. as the end-effector. 


\fg{end-effector}{end-effector}{1}{End-effector can be on different part of the object: static part of spatula (a) or end of the transformable part of Minion piggybank (b)}

Below we detail two key steps for Romeo to generate a robotic arm from the selected part: \one segmenting the transformable part and \two generating joints between segmented links. 



\subsection{Segmentation of the Transformable Part}
Romeo employs two ways of segmentation based on the shape of the transformable part: slender and non-slender.

\insec{Slender shape}
Romeo considers a transformable part as slender if one of the dimensions (X/Y/Z) is at least four times as longer as the others \footnote{Recall that we have preprocessed the object to make it as axis-aligned as possible.}. 
For this type of shape, Romeo segments it by quartering along its longest axis into links of equal lengths (\fgref{segment}a). 

\insec{Non-slender shape}
For a non-slender transformable part, we need to first determine a \textit{principal axis}, the normal axis of the plane whereon the transformable part unfolds. Romeo picks the cross-section of transformable part with the largest area
as the unfolding plane and set its normal as the principal axis. Then Romeo quarters the transformable part equally around the principal axis to operate segmentation (\fgref{segment}b).
\fg{segment}{segment}{1}{Two types of segmentation, (a) slender (b) non-slender shapes.
}

\subsection{Searching for Robotic Arm Configurations}
Romeo generates a four-joint robotic arm, 
divided into two groups---one \textit{steering} joint and three \textit{driving} joints (\fgref{joints}). 
Generally speaking, a steering joint controls the overall orientation of the robotic arm, the choice of which is based on the motion points; the driving joints control how far the joint can reach, parallel to the principal axis. 
By only changing the axis of the steering joint while keeping the same design of the driving joints, Romeo simplifies the number of possible configurations to three. Then for each configuration, Romeo samples value at each joint within its range to calculate the workspace and compare with each other to pick the workspace closest to the motion points.



\fg{dh}{DHParameter}{1}{Four parameters to translate the coordinates. The figure is borrowed from Wikipedia: DH Parameters (Accessed 5/5/2020)
}

To compute the workspace, we first introduce modified Denavit–Hartenberg (DH) parameters \cite{hartenberg1955kinematic}--- four parameters ($a$, $\alpha$, $d$, $\theta$) in mechanical engineering to represent spatial linkage systems, such as a robotic arm (\fgref{dh} indexes each parameter). With DH parameters, the position and orientation of the $i^\text{th}$ joint relative to the $i$-1$^\text{th}$ joint can be represented as a transformation matrix ($c$ for $cos$, $s$ for $sin$):
\begin{equation*}
    \prescript{i-1}{i}{T} = 
    \begin{bmatrix}
    c\theta_i & -s\theta_i & 0 & a_{i-1}\\
    s\theta_ic\alpha_{i-1} & c\theta_ic\alpha_{i-1} & -s\alpha_{i-1} & -s\alpha_{i-1}d_i \\
    s\theta_is\alpha_{i-1} & c\theta_is\alpha_{i-1} & c\alpha_{i-1} & c\alpha_{i-1}d_i \\
    0 & 0 & 0 & 1 \\
    \end{bmatrix}
\end{equation*}
Therefore, the cartesian position and orientation of the end-effector can be computed by $^0_NT$ using forward kinematics:
\begin{equation*}
    \prescript{0}{N}{T}=\prescript{0}{1}{T}\prescript{1}{2}{T}\prescript{2}{3}{T}...\prescript{N-1}{N}{T}
\end{equation*}


In our case, upon the transformable part selection, Romeo first generates a robotic arm with seven joints representing the three configurations in a single form of a robotic arm. Three joints correspond to the steering joint for each configuration, while another three joints represent the driving joints. The $7^{th}$ joint represents the end-effector. 
With the list of DH parameters, Romeo computes the position and orientation of the end-effector by sampling values at each joint. For each configuration, Romeo only changes the value of the corresponding steering joint while keeping the other two constant.

\fg{workspace}{workspace}{1}{Visualized workspace (a). Double clicking makes the motion points outside, highlighted in red (b), snaps to the closet workspace (c). 
}

To determine the best configuration for the user-defined task, Romeo takes the sampled end-effector positions and orientations as a workspace (represented as its convex hull as depicted in \fgref{workspace}) and calculates the minimal distance to the motion points. If there exists the orientation requirement, Romeo selects end-effector position only within the specified orientation range. 
Finally, Romeo picks the configuration whose distance to the motion points has the minimal RMSE (Root-Mean-Square-Deviation).
In the following, we discuss details about arranging the steering and driving joints in distinct cases based on the resultant robotic arm: 

\fg{joints}{joints}{1}{Corresponding robotic arm representatives and the joint placement for three distinct cases.}

\one \insec{Unfolded robotic arm}
For object in a slender shape, the generated embedded robotic arm is already unfolded at its initial state, \eg spatula (\fgref{joints}a). Therefore, Romeo places the steering joint at the location which is farthest to the end-effector (\fgref{joints}b). 

\two \insec{Folded robotic arm/End-effector on transformable part}
For objects that have a non-slender shape and the end-effector is on the transformable part, the initial state of the embedded robotic arm is folded, \eg a piggybank (\fgref{joints}c). Romeo places the steering joint at the second joint from the static part (\fgref{joints}d), because if transformable part locates in the middle of two static parts, it is likely cause collision with the object when unfolding, due to the steering joint placed at the first joint (\fgref{collision}). 

\fg{collision}{collision}{1}{
 If the steering joint locates at the first joint while the transformable part lies in the middle, both cases of b and c will cause collision between geometries.
}


\three \insec{Folded robotic arm/End-effector on static part}
For objects with a non-slender shape but the end-effector is not in the selected part (\eg See stamp in  \fgref{joints}e), the initial state of the embedded robotic arm is set to the folded state. 
Romeo places the steering joint to where it is farthest from the static part when the transformable part is unfolded. The steering joint will be fixed to the ground, so to become a base of the generated arm (\fgref{joints}f). 

Please refer to Appendix for the complete DH tables. 
{\bf Other search criteria} include the following.

\insec{Matching the orientation}
When searching for the best configuration, we also need to match the orientation if it is specified at a certain motion point. Romeo uses the X-axis of the end-effector joint as it matches the pointing direction. 
The pointing direction of the stamp is represented by the $X_7$ axis (as illustrated as a red arrow in \fgref{orientation}a), which is used to match the user-specified orientation at the motion point (black arrow in \fgref{orientation}b).
Recall the transformation matrix of end-effector relative to the base $\prescript{0}{N}{T}$:
\begin{equation*}
    \prescript{0}{N}{T} = 
	\left[
	\begin{array}{ccc|c}
	n & o & a & t \\ \hline
	0 & 0 & 0 & 1\\
	\end{array}
	\right]
\end{equation*}
in which $n$ is a normalized vector ($n\in R^3$) representing the direction of X-axis of the end-effector. Take the users specify direction as $dir$ and the positions of the end-effector whose orientation matches $|n-dir|\leq 0.5$ are included to compute the minimal distance to the motion points.
\fg{orientation}{orientation}{1}{Matching the orientation by aligning the x-axis of the end-effector with the arrow direction of the spherical widget control. The coordinate system represent the target orientation of the end-effector (a). Black arrow indicates the specified orientation (b).}

\insec{Obstacle avoidance}
The collision between the transformed robotic arm and the remaining object itself (mainly the static part) is a significant factor that determines the performance of the robotic arm. Currently, Romeo defines the bounding box of the static part(s) as an obstacle. While searching for the best configuration, Romeo eliminates pose options of the robotic arm in which a joint position intersects the obstacle's bounding box.

\insec{Users' modification}
After the configuration of the robotic arm is determined, there could be cases where some of the motion points still situate outside of the workspace. To address this, Romeo detects if the point is inside the workspace by forming a convex hull of all the sample points and highlighting exterior points in red (\fgref{workspace}b-c). 
Then Romeo enables users to double click on such points, which snaps it to the nearest points on the workspace surface. 



\section{\#4 Generating Components for Fabrication}
As the final step, Romeo automatically generates 3D printable components. 
To generate assembly-ready components, Romeo performs a series of Boolean operations to create space for motor fasteners, joint connectors, screw holes and end-effectors. 

\textbf{Motor fasteners} \ Romeo allows users to choose whether to embed motors when generating the embedded arm. 
For manual transformation, Romeo generates hinges between each link. For motors integration, the system computes each motor's pivot position. 
The links connect directly to the motors without any power transmission system (\eg gear system, four-bar-linkage system).
Therefore, pivot positions are set in a way that it prevents links from colliding with each other, as shown in \fgref{pivot}a. 
Based on the pivot position, Romeo adds a `shell' to mount the motor.
\fg{pivot}{pivot_and_connector}{1}{Two types of joint connectors.
}

\textbf{Joint connectors} \  Romeo creates two types of joint connectors. Type A (\fgref{pivot}a) is used when the two connecting links move in the same plane, while type B (\fgref{pivot}b) is used when there is an offset between the two moving plane. 
For type A
, the system fillets on the link corners to prevent self-collision and to secure spaces for the joint connectors to move. 
For type B, Romeo creates a fastening mechanism using screws. 

\textbf{Screw holes} \ 
Screws are needed to fasten motors and joints with the links, thus, Romeos adds screw holes as shown in \fgref{gripper}a. The system creates rivet holes specifically for Dynamixel XL-320 motors\footnote{\url{http://www.robotis.us/dynamixel-xl-320/}}, although other types of motors can be supported in future work.

\fg{gripper}{gripper}{1}{Romeo generates screw holes (a) and gripper (b).}
\vspace{-2.0px}

\textbf{End-effectors} \ 
For motion points that include `picking and placing' action, Romeo generates a robotic gripper
as the end-effector which will need additional motor to actuate it. 
We currently provide one universal gripper design that can be directly imported into Romeo (\fgref{gripper}b).
For `attaching to a surface' action, Romeo employs AutoConnect's approach \cite{koyama2015autoconnect} to generate clamp fasteners to be associated with the different types of attaching surface.






\section{FABRICATION AND SOFTWARE IMPLEMENTATION}
Romeo's front end is written in JavaScript using jQuery 
for UI development, three.js 
for 3D graphics, and ThreeCSG
for progressively generating the geometry of components. The back end is written in MATLAB 2018a. Everything runs on a MacBook Pro (15-inch, 2016 year) with a 2.7 GHz Intel i7 and 16GB 2133 MHz LPDDR3 memory. In our design session and demonstrations, the interface runs on a Google Chrome web browser. For remote participants, we use TeamViewer \footnote{\url{https://www.teamviewer.com/en-us/}} for screen sharing and remote control. We use Dynamixel XL-320 motors to actuate the transformed objects, all parts are 3D printed by the Ultimaker S5 using white PLA.

\section{Examples Generated using Romeo}

We present a series of examples created by Romeo.
As we focused on demonstrating Romeo's potential to augment objects' functionality, we used other CAD tools for non transformable-related postprocessing, \eg making a fastener to affix a spatula to the side of a pot, adding a small water tank to the flowerpot for keeping water. 

Figure 17---23 showcase exemplary augmentation of daily objects by transforming part of the objects into a robotic arm: A cup holder's lower half can be (manually) extended to attach to a chair to free the user's hands (\fgref{cupholder}). A paper towel stand unfolds its base to hold the paper towel helping user pull apart a sheet (thus avoid touching the roll using wet hands) (\fgref{towel}). A spatula can be robotized so that it is affixed to the pot's handle and to automatically stir the pot to free a user's hands (\fgref{spatula}). A conventional stamp with an UIST 2020 logo can be robotized to become a self-inking stamp (\fgref{stamp}). For a busy office worker who is likely to forget to water the plant, the flowerpot can be transformed to water itself regularly (\fgref{flowerpot}). Under the contagious COVID-19 pandemic, a box of sanitizing wipes can disinfect the doorknob every time someone touches the doorknob (\fgref{tissue}). 


\fgref{fig1}d and \fgref{piggybank} depict examples with different 3D models transformed into a `stealing piggybank' using Romeo. 

\fgh{towel}{paper_towel}{1}{By robotizing a paper towel holder, a user can pull one sheet without contaminating the others.}

\fgh{stamp}{stamp}{1}{A conventional stamp can be robotized by Romeo to become a self-inking stamp.
}

\fgh{spatula}{spatula}{1}{Romeo can robotize a spatula to automatically stir the pot when unattended.}
\fgh{tissue}{tissue}{1}{A tissue box can be stick to the door and wipe the doorknob every time someone opens the door.
}
\fgh{cupholder}{cupholder}{1}{A cup holder can extend the lower part of it to attach to the chair.}

\fgh{flowerpot}{flowerpot}{1}{A flowerpot can be robotized to automatically water the plant according to a predefined schedule.
}

\fgh{piggybank}{piggybank}{1}{Different 3D objects can be augmented into a stealing coin piggy bank using Romeo. The examples are made by the participants in the design sessions: a dodo bird (a) and a Tesla cybertruck (b)}

\section{Design Sessions}
We conducted informal, qualitative design sessions 
to validate Romeo's ability to support non-expert users' design of transformables using existing 3D models. 

\subsection{Participants}
We recruited eight participants (aged 23-26, female=5, male=3).
No participant had background in Mechanical Engineering. Five participants had Electrical and Computer Engineering background, two of which had experience in the computer vision area of Robotics. The rest Participants did not have background in engineering. Amongst all participants, three had experience using CAD tools before.

\subsection{Apparatus, Tasks and Procedure}
We split design tasks in two sessions spanning two days to budget time for 3D printing the resulting transformables. In the first session, participants designed transformables via TeamViewer's remote desktop to interact with Romeo running on the experimenter's computer. The components were then 3D printed for assembly, tested with its designed task in the second session on the next day. 
However, due to the COVID-19 outbreaks, only four participants were able to meet in person to join the second part for assembly and test the fabricated results. Therefore, alternatively, the results of remote-only participants' designs are assembled and tested by the experimenter and the process is recorded to be shown to these participants. 

\insec{Design Session (Day 1)}
To start, each participant was introduced some basic knowledge of how an robotic arm works (\eg how a 6-DOF arm moves along with rotary joints) and how to use Romeo by walking through an example of making a 
snowman-shaped coin-stealing piggybank. 
Then participants proceeded to use Romeo for one controlled and one open-design tasks.
Each task's objective is using Romeo to augment an object by creating transformable parts to add a new functionality. 
In the first (controlled) task, the participant was to replicate either the spatula (\fgref{spatula}) or the tissue box examples (\fgref{tissue}). 
Specifically, the spatula would be transformed to stir in a pot and the tissue box to pick a tissue and wipe a door knob.
In the second open-design task, the participant used a 3D object of their own choice to design a coin-stealing piggybank.

After the first session, we manually post-processed the resulting components, creating holes for cables and hollowing internal volume to reduce printing time.

\insec{Assembly \& Interaction Session (Day 2)} 
For the four participants who were able to join the in-person assembly session, they were given instructions (\fgref{instruction}b) based on which they assembled the printed parts. The logged data of their design were fed into a universal code for actuating the motors. Then the participants test the automated tasks, if they match with the simulated animation.


\fg{instruction}{instruction}{1}{A participant installs the printed components (a) following a provided instruction (b).}

\insec{Metrics \& Measurements}
At the end of the Day 1 session, participants filled out a questionnaire regarding overall user experience, including the difficulty to learn how to use Romeo and whether the process required extra knowledge than what they had expected (in Likert scale of 1-7). 
After the Day 2 session, the four participants filled out another questionnaire to answer how difficult it was to assemble the fabricated results, whether the installed mechanisms behave as they expected, and perceived usefulness of having such augmented transformable objects. 
Finally, we conducted a brief interview and solicited feedback about the entire process and suggestions.

\subsection{Results}
Overall, during the first session of using Romeo to create transformable designs, participants were able to complete the controlled and open-ended tasks. 
During the second session with four participants, they were able to complete assembly given the instruction, linking parts in order and connect motors in the right orientation.
They were mostly satisfied with the design flow and underlying automation that aided the design and assembly (Satisfaction score in Mean=5.25---6.75/7 across questions).
See Figure~\ref{fig:study-quant} in Appendix for quantitative summary of the result and Figure~\ref{fig:study} for all participants design.




\subsection{Observations \& Findings}
We present the findings 
by questionnaire responses, analyzing the logged observation of participants' behaviors and spoken responses during the design session as follows.


Users enjoyed the simplicity of Romeo's design process with automation, noting that it is very simple with only four steps (P8).
Participants seemed to highly value the automated mechanism generation, as it removed the needs in engineering expertise (P4) and reduced the required intellectual efforts (P3)---``After I specify points, Romeo helps me do the other things even though I don't have any related background'' (P1).
``[There was] No need to consider about the transformation and movement design. All I need is a clear goal'' (P4).
Although Romeo requires some manual adjustment if the design is not viable (\eg double-clicking an outside point to snap it back to the workspace)
participants felt the overall process is ``completely automatic'' (P6-7).
Participants also appreciated the animation function that helps visualize and validate resulting movement, showing their satisfaction about the simulation of a designed task.``It's very clear to help me understand how the movement is achieved'' (P1).

The major challenge seems to be understanding where to place the end-effector.
As introduced in the earlier section,
users can place the end-effector either on transformable part (as in the piggy-bank model, \fgref{piggybank}) or on the static part (as in the spatula example, \fgref{spatula}), but
some users have difficulties in determining which of these two approaches to use.
%
``
I can try either one to decide which is right'' (P1). 

Participants also had difficulties in understanding how a transformable will be anchored, \ie where the base is.
For example, P1 selected only the middle part of the spatula handle in the first trial, thinking she needs to leave the top part to serve as the base that anchors the spatula while stirring.
Another participant also questioned ``If I specify the middle part, will the top part move with it?'' (P8).
However, after seeing the generated results, participants naturally understood the consequence, becoming able to select the placement of end-effector in the subsequent tasks:
``I did not understand why we did them. I understood it after couple of examples though!'' (P2).

Such confusion was partial result of our design choice, abstracting away low-level details for users, while allowing them to rapidly explore different designs and see results, so to iterate on their earlier decision (\eg where to place the base/end-effector). 
We can also support users to explore different possibilities simultaneously to choose between different results. 






\section{Discussion, Limitations \& Future Work}

\textbf{Scale of Mechanism} \
In order to embed motors, Romeo currently requires the input 3D models have sufficient volume to house the motors in each link. 
Further, the scale of an input 3D model also determines the possible range of movement of a robotic arm transformed from part of the object.
As a next step, Romeo can support automtically scaling a 3D model to meet both the requirements of housing motors and accommodating sufficiently large workspace, while estimating the printing time/cost to help users make an informed decision whether to use the scaled model.
selects too small parts to embed transformable mechanisms, as also questioned by one of our participant: ``If I only want to transform ears, will that be too small?'' (P7). 
Guiding users with alert message to notify that the parts need to be larger, and automatically adjusting to the minimal-scale would aid users to avoid any potential design errors.

\textbf{Challenges in Selecting a Transformable Part} 
To simplify the design process, we currently allow users to specify a transformable part by sweeping an object's cross section along one of X/Y/Z axis.
Future work should explore techniques for more expressive selection of transformable parts in order to support cases such as selecting the face of a bear, the wing of a bird, the arm of a minion character.
One possible solution is to include functions provided in existing CAD tool such as a brushing interface to select parts by painting areas of interest and automatically offsetting enough volumes \cite{MeshMixerSeg} or automating complex mesh-segmentation with semantic segmentation \cite{3DMeshSegmentation}.
Further, users' selection of transformable part may affect the object's original functionalities. Future work may enable highlighting the original functional part of the object using a machine learning model to guide users not to select that part as transformable. 

\textbf{Desired Task Specification} 
Assigning task by each point appears unintuitive to some users.
P4 asked 
``Do I need to specify the action for every point?'', and
P8 thought there would be some high-level task description to select, questioning ``Is the following a trajectory action means the action of wiping?''.
Future work can provide a user with an option to specify the task type upfront and provide more details on demand (\eg at which point to pick/place) and real-time animation can also provide users with more intuitive feedback.
Another interesting possibility is allowing users to describe a high-level task and search in a library of existing transformable designs that meet such requirements. 

\textbf{Pre/post-processing}
Currently, an input 3D model needs to be pre-processed to simplify meshes, to align the bounding box's axis parallel to the coordinates, to rotate and locate it in a way that motion points can match the world-coordinate to aid the user's understanding of its relative position to the target 3D object.
To reduce weight, material cost and printing time, one common post-processing step is hollowing the components using existing 3D modelling tool. 
In the future, both pre- and post- processing steps can be engineered into Romeo for better integration.


\textbf{Error Detection and Automatic Geometry Processing}
Current segmentation algorithm might cause disjoint links if the transformable part has high concavity such as branching shapes. Future work can detect the concavity of selected transformable part and explore an adaptive segmentation method to avoid creating disjoint components. 
Romeo automatically generates the motion path for conducting the tasks, however, without detecting the collision between the motion points, which may result in self collision as also observed in one case in our assembly session. An interesting future direction could be applying trajectory planning algorithm and obstacle avoidance of robotic arm to generate collision-free motion path.

\section{Acknowledgement}
We thank the reviewers for their valuable feedback. We also thank participants for participating in the user study. This work was funded in part by Adobe.

\balance{}

\bibliographystyle{SIGCHI-Reference-Format}
\bibliography{main}


\begin{thebibliography}{00}


\ifx \showCODEN    \undefined \def \showCODEN     #1{\unskip}     \fi
\ifx \showDOI      \undefined \def \showDOI       #1{{\tt DOI:}\penalty0{#1}\ }
  \fi
\ifx \showISBNx    \undefined \def \showISBNx     #1{\unskip}     \fi
\ifx \showISBNxiii \undefined \def \showISBNxiii  #1{\unskip}     \fi
\ifx \showISSN     \undefined \def \showISSN      #1{\unskip}     \fi
\ifx \showLCCN     \undefined \def \showLCCN      #1{\unskip}     \fi
\ifx \shownote     \undefined \def \shownote      #1{#1}          \fi
\ifx \showarticletitle \undefined \def \showarticletitle #1{#1}   \fi
\ifx \showURL      \undefined \def \showURL       #1{#1}          \fi

\bibitem{bacher2012fabricating}
{Moritz B{\"a}cher}, {Bernd Bickel}, {Doug~L James}, {and} {Hanspeter Pfister}.
  2012.
\newblock \showarticletitle{Fabricating articulated characters from skinned
  meshes}.
\newblock {\em ACM Transactions on Graphics (TOG)\/} {31}, 4 (2012), 1--9.
\newblock


\bibitem{bacher2015linkedit}
{Moritz B{\"a}cher}, {Stelian Coros}, {and} {Bernhard Thomaszewski}. 2015.
\newblock \showarticletitle{LinkEdit: interactive linkage editing using
  symbolic kinematics}.
\newblock {\em ACM Transactions on Graphics (TOG)\/} {34}, 4 (2015), 1--8.
\newblock


\bibitem{cali20123d}
{Jacques Cal{\`\i}}, {Dan~A Calian}, {Cristina Amati}, {Rebecca Kleinberger},
  {Anthony Steed}, {Jan Kautz}, {and} {Tim Weyrich}. 2012.
\newblock \showarticletitle{3D-printing of non-assembly, articulated models}.
\newblock {\em ACM Transactions on Graphics (TOG)\/} {31}, 6 (2012), 1--8.
\newblock


\bibitem{chen2016reprise}
{Xiang'Anthony' Chen}, {Jeeeun Kim}, {Jennifer Mankoff}, {Tovi Grossman},
  {Stelian Coros}, {and} {Scott~E Hudson}. 2016.
\newblock \showarticletitle{Reprise: A design tool for specifying, generating,
  and customizing 3D printable adaptations on everyday objects}. In {\em
  Proceedings of the 29th Annual Symposium on User Interface Software and
  Technology}. 29--39.
\newblock


\bibitem{chen2018forte}
{Xiang'Anthony' Chen}, {Ye Tao}, {Guanyun Wang}, {Runchang Kang}, {Tovi
  Grossman}, {Stelian Coros}, {and} {Scott~E Hudson}. 2018.
\newblock \showarticletitle{Forte: User-driven generative design}. In {\em
  Proceedings of the 2018 CHI Conference on Human Factors in Computing
  Systems}. 1--12.
\newblock


\bibitem{coros2013computational}
{Stelian Coros}, {Bernhard Thomaszewski}, {Gioacchino Noris}, {Shinjiro Sueda},
  {Moira Forberg}, {Robert~W Sumner}, {Wojciech Matusik}, {and} {Bernd Bickel}.
  2013.
\newblock \showarticletitle{Computational design of mechanical characters}.
\newblock {\em ACM Transactions on Graphics (TOG)\/} {32}, 4 (2013), 1--12.
\newblock


\bibitem{davidoff2011mechanical}
{Scott Davidoff}, {Nicolas Villar}, {Alex~S Taylor}, {and} {Shahram Izadi}.
  2011.
\newblock \showarticletitle{Mechanical hijacking: how robots can accelerate
  UbiComp deployments}. In {\em Proceedings of the 13th international
  conference on Ubiquitous computing}. ACM, 267--270.
\newblock


\bibitem{desai2019geppetto}
{Ruta Desai}, {Fraser Anderson}, {Justin Matejka}, {Stelian Coros}, {James
  McCann}, {George Fitzmaurice}, {and} {Tovi Grossman}. 2019.
\newblock \showarticletitle{Geppetto: Enabling Semantic Design of Expressive
  Robot Behaviors}. In {\em Proceedings of the 2019 CHI Conference on Human
  Factors in Computing Systems}. 1--14.
\newblock


\bibitem{desai2018automatic}
{Ruta Desai}, {Margarita Safonova}, {Katharina Muelling}, {and} {Stelian
  Coros}. 2018.
\newblock \showarticletitle{Automatic design of task-specific robotic arms}.
\newblock {\em arXiv preprint arXiv:1806.07419\/} (2018).
\newblock


\bibitem{desai2017computational}
{Ruta Desai}, {Ye Yuan}, {and} {Stelian Coros}. 2017.
\newblock \showarticletitle{Computational abstractions for interactive design
  of robotic devices}. In {\em 2017 IEEE International Conference on Robotics
  and Automation (ICRA)}. IEEE, 1196--1203.
\newblock


\bibitem{garg2016computational}
{Akash Garg}, {Alec Jacobson}, {and} {Eitan Grinspun}. 2016.
\newblock \showarticletitle{Computational design of reconfigurables.}
\newblock {\em ACM Trans. Graph.\/} {35}, 4 (2016), 90--1.
\newblock


\bibitem{ha2018computational}
{Sehoon Ha}, {Stelian Coros}, {Alexander Alspach}, {James~M Bern}, {Joohyung
  Kim}, {and} {Katsu Yamane}. 2018.
\newblock \showarticletitle{Computational design of robotic devices from
  high-level motion specifications}.
\newblock {\em IEEE Transactions on Robotics\/} {34}, 5 (2018), 1240--1251.
\newblock


\bibitem{hartenberg1955kinematic}
{Richard~S Hartenberg} {and} {Jacques Denavit}. 1955.
\newblock \showarticletitle{A kinematic notation for lower pair mechanisms
  based on matrices}.
\newblock  (1955).
\newblock


\bibitem{HUANG2016127}
{Yi-Jheng Huang}, {Shu-Yuan Chan}, {Wen-Chieh Lin}, {and} {Shan-Yu Chuang}.
  2016.
\newblock \showarticletitle{Making and animating transformable 3D models}.
\newblock {\em Computers \& Graphics\/}  {54} (2016), 127 -- 134.
\newblock
\showISSN{0097-8493}
\showDOI{%
\url{http://dx.doi.org/https://doi.org/10.1016/j.cag.2015.07.014}}
\newblock
\shownote{Special Issue on CAD/Graphics 2015.}


\bibitem{hudson2016understanding}
{Nathaniel Hudson}, {Celena Alcock}, {and} {Parmit~K Chilana}. 2016.
\newblock \showarticletitle{Understanding newcomers to 3D printing:
  Motivations, workflows, and barriers of casual makers}. In {\em Proceedings
  of the 2016 CHI Conference on Human Factors in Computing Systems}. 384--396.
\newblock


\bibitem{3DMeshSegmentation}
{Evangelos Kalogerakis}, {Aaron Hertzmann}, {and} {Karan Singh}. 2010.
\newblock \showarticletitle{Learning 3D Mesh Segmentation and Labeling}.
\newblock {\em ACM Trans. Graph.\/} {29}, 4, Article Article 102 (July 2010),
  12 pages.
\newblock
\showISSN{0730-0301}
\showDOI{%
\url{http://dx.doi.org/10.1145/1778765.1778839}}


\bibitem{kazi2017dreamsketch}
{Rubaiat~Habib Kazi}, {Tovi Grossman}, {Hyunmin Cheong}, {Ali Hashemi}, {and}
  {George~W Fitzmaurice}. 2017.
\newblock \showarticletitle{DreamSketch: Early Stage 3D Design Explorations
  with Sketching and Generative Design.}. In {\em UIST}, Vol.~14. 401--414.
\newblock


\bibitem{kovacs2018trussformer}
{Robert Kovacs}, {Alexandra Ion}, {Pedro Lopes}, {Tim Oesterreich}, {Johannes
  Filter}, {Philipp Otto}, {Tobias Arndt}, {Nico Ring}, {Melvin Witte}, {Anton
  Synytsia}, {and} {others}. 2018.
\newblock \showarticletitle{Trussformer: 3d printing large kinetic structures}.
  In {\em Proceedings of the 31st Annual ACM Symposium on User Interface
  Software and Technology}. 113--125.
\newblock


\bibitem{koyama2015autoconnect}
{Yuki Koyama}, {Shinjiro Sueda}, {Emma Steinhardt}, {Takeo Igarashi}, {Ariel
  Shamir}, {and} {Wojciech Matusik}. 2015.
\newblock \showarticletitle{AutoConnect: computational design of 3D-printable
  connectors}.
\newblock {\em ACM Transactions on Graphics (TOG)\/} {34}, 6 (2015), 1--11.
\newblock


\bibitem{li2015foldabilizing}
{Honghua Li}, {Ruizhen Hu}, {Ibraheem Alhashim}, {and} {Hao Zhang}. 2015.
\newblock \showarticletitle{Foldabilizing furniture.}
\newblock {\em ACM Trans. Graph.\/} {34}, 4 (2015), 90--1.
\newblock


\bibitem{li2019robiot}
{Jiahao Li}, {Jeeeun Kim}, {and} {Xiang'Anthony' Chen}. 2019a.
\newblock \showarticletitle{Robiot: A Design Tool for Actuating Everyday
  Objects with Automatically Generated 3D Printable Mechanisms}. In {\em
  Proceedings of the 32nd Annual ACM Symposium on User Interface Software and
  Technology}. 673--685.
\newblock


\bibitem{DBLP:conf/uist/LiKC19}
{Jiahao Li}, {Jeeeun Kim}, {and} {Xiang~'Anthony' Chen}. 2019b.
\newblock \showarticletitle{Robiot: {A} Design Tool for Actuating Everyday
  Objects with Automatically Generated 3D Printable Mechanisms}. In {\em
  Proceedings of the 32nd Annual {ACM} Symposium on User Interface Software and
  Technology, {UIST} 2019, New Orleans, LA, USA, October 20-23, 2019}.
  673--685.
\newblock
\showDOI{%
\url{http://dx.doi.org/10.1145/3332165.3347894}}


\bibitem{megaro2015interactive}
{Vittorio Megaro}, {Bernhard Thomaszewski}, {Maurizio Nitti}, {Otmar Hilliges},
  {Markus Gross}, {and} {Stelian Coros}. 2015.
\newblock \showarticletitle{Interactive design of 3D-printable robotic
  creatures}.
\newblock {\em ACM Transactions on Graphics (TOG)\/} {34}, 6 (2015), 1--9.
\newblock


\bibitem{mehta2016design}
{Ankur Mehta}, {Nicola Bezzo}, {Peter Gebhard}, {Byoungkwon An}, {Vijay Kumar},
  {Insup Lee}, {and} {Daniela Rus}. 2016.
\newblock \showarticletitle{A design environment for the rapid specification
  and fabrication of printable robots}. In {\em Experimental Robotics}.
  Springer, 435--449.
\newblock


\bibitem{mehta2015integrated}
{Ankur Mehta}, {Joseph DelPreto}, {and} {Daniela Rus}. 2015.
\newblock \showarticletitle{Integrated codesign of printable robots}.
\newblock {\em Journal of Mechanisms and Robotics\/} {7}, 2 (2015).
\newblock


\bibitem{mehta2018robot}
{Ankur~M Mehta}, {Joseph DelPreto}, {Kai~Weng Wong}, {Scott Hamill}, {Hadas
  Kress-Gazit}, {and} {Daniela Rus}. 2018.
\newblock \showarticletitle{Robot creation from functional specifications}.
\newblock In {\em Robotics Research}. Springer, 631--648.
\newblock


\bibitem{ramakers2016retrofab}
{Raf Ramakers}, {Fraser Anderson}, {Tovi Grossman}, {and} {George Fitzmaurice}.
  2016.
\newblock \showarticletitle{Retrofab: A design tool for retrofitting physical
  interfaces using actuators, sensors and 3d printing}. In {\em Proceedings of
  the 2016 CHI Conference on Human Factors in Computing Systems}. 409--419.
\newblock


\bibitem{roumen2018grafter}
{Thijs~Jan Roumen}, {Willi M{\"u}ller}, {and} {Patrick Baudisch}. 2018.
\newblock \showarticletitle{Grafter: Remixing 3D-printed machines}. In {\em
  Proceedings of the 2018 CHI Conference on Human Factors in Computing
  Systems}. 1--12.
\newblock


\bibitem{schulz2017interactive}
{Adriana Schulz}, {Cynthia Sung}, {Andrew Spielberg}, {Wei Zhao}, {Robin
  Cheng}, {Eitan Grinspun}, {Daniela Rus}, {and} {Wojciech Matusik}. 2017.
\newblock \showarticletitle{Interactive robogami: An end-to-end system for
  design of robots with ground locomotion}.
\newblock {\em The International Journal of Robotics Research\/} {36}, 10
  (2017), 1131--1147.
\newblock


\bibitem{teibrich2015patching}
{Alexander Teibrich}, {Stefanie Mueller}, {Fran{\c{c}}ois Guimbreti{\`e}re},
  {Robert Kovacs}, {Stefan Neubert}, {and} {Patrick Baudisch}. 2015.
\newblock \showarticletitle{Patching physical objects}. In {\em Proceedings of
  the 28th Annual ACM Symposium on User Interface Software \& Technology}.
  83--91.
\newblock


\bibitem{thomaszewski2014computational}
{Bernhard Thomaszewski}, {Stelian Coros}, {Damien Gauge}, {Vittorio Megaro},
  {Eitan Grinspun}, {and} {Markus Gross}. 2014.
\newblock \showarticletitle{Computational design of linkage-based characters}.
\newblock {\em ACM Transactions on Graphics (TOG)\/} {33}, 4 (2014), 1--9.
\newblock


\bibitem{ureta2016interactive}
{Francisca~Gil Ureta}, {Chelsea Tymms}, {and} {Denis Zorin}. 2016.
\newblock \showarticletitle{Interactive modeling of mechanical objects}. In
  {\em Computer Graphics Forum}, Vol.~35. Wiley Online Library, 145--155.
\newblock


\bibitem{MeshMixerSeg}
{Jonny Yeu}. 2019.
\newblock Convert Single-Color STL files into Multi-Body Models using
  MeshMixer.
\newblock   (2019).
\newblock
\showURL{%
\url{https://support.mosaicmfg.com/hc/en-us/articles/115002827014}}


\bibitem{yu2017computational}
{Christopher Yu}, {Keenan Crane}, {and} {Stelian Coros}. 2017.
\newblock \showarticletitle{Computational design of telescoping structures}.
\newblock {\em ACM Transactions on Graphics (TOG)\/} {36}, 4 (2017), 1--9.
\newblock


\bibitem{yuan2018computational}
{Ye Yuan}, {Changxi Zheng}, {and} {Stelian Coros}. 2018.
\newblock \showarticletitle{Computational design of transformables}. In {\em
  Computer Graphics Forum}, Vol.~37. Wiley Online Library, 103--113.
\newblock


\bibitem{Boxelization}
{Yahan Zhou}, {Shinjiro Sueda}, {Wojciech Matusik}, {and} {Ariel Shamir}. 2014.
\newblock \showarticletitle{Boxelization: Folding 3D Objects into Boxes}.
\newblock {\em ACM Trans. Graph.\/} {33}, 4, Article Article 71 (July 2014), 8
  pages.
\newblock
\showISSN{0730-0301}
\showDOI{%
\url{http://dx.doi.org/10.1145/2601097.2601173}}


\end{thebibliography}

\clearpage
\newpage

\onecolumn
\section{Appendix}
\subsubsection{DH Parameter Table}
\begin{figure}[h]
    \centering
    \includegraphics[width=\columnwidth]{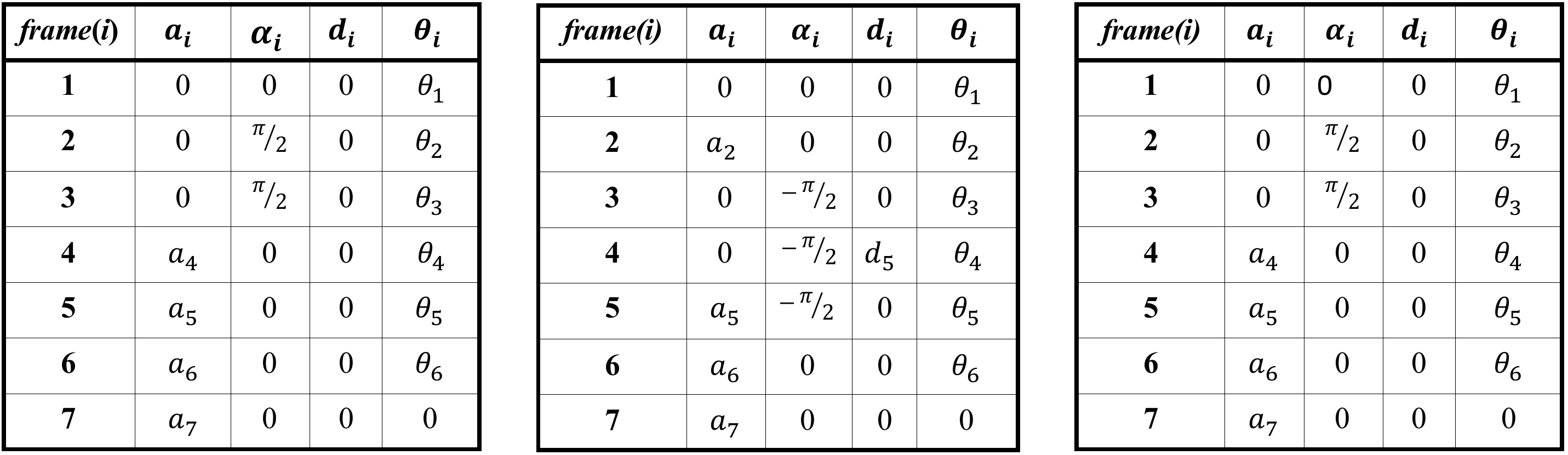}
    \caption{Complete set of DH parameters list for Figure12 examples. Each table corresponds to (from left to right) Spatula with unfolded robotic arm in the initial state (a-b), Minion piggy-bank with folded robotic arm in the initial state (c-d), and Stamp with folded robotic arm but the end-effector on static part (e-f)}
    \label{fig:dh_param_tables}
\end{figure}


\subsubsection{Design Session Quantitative Results and Design Results}

\begin{figure}[h]
  \centering
  \includegraphics[width=0.9\columnwidth]{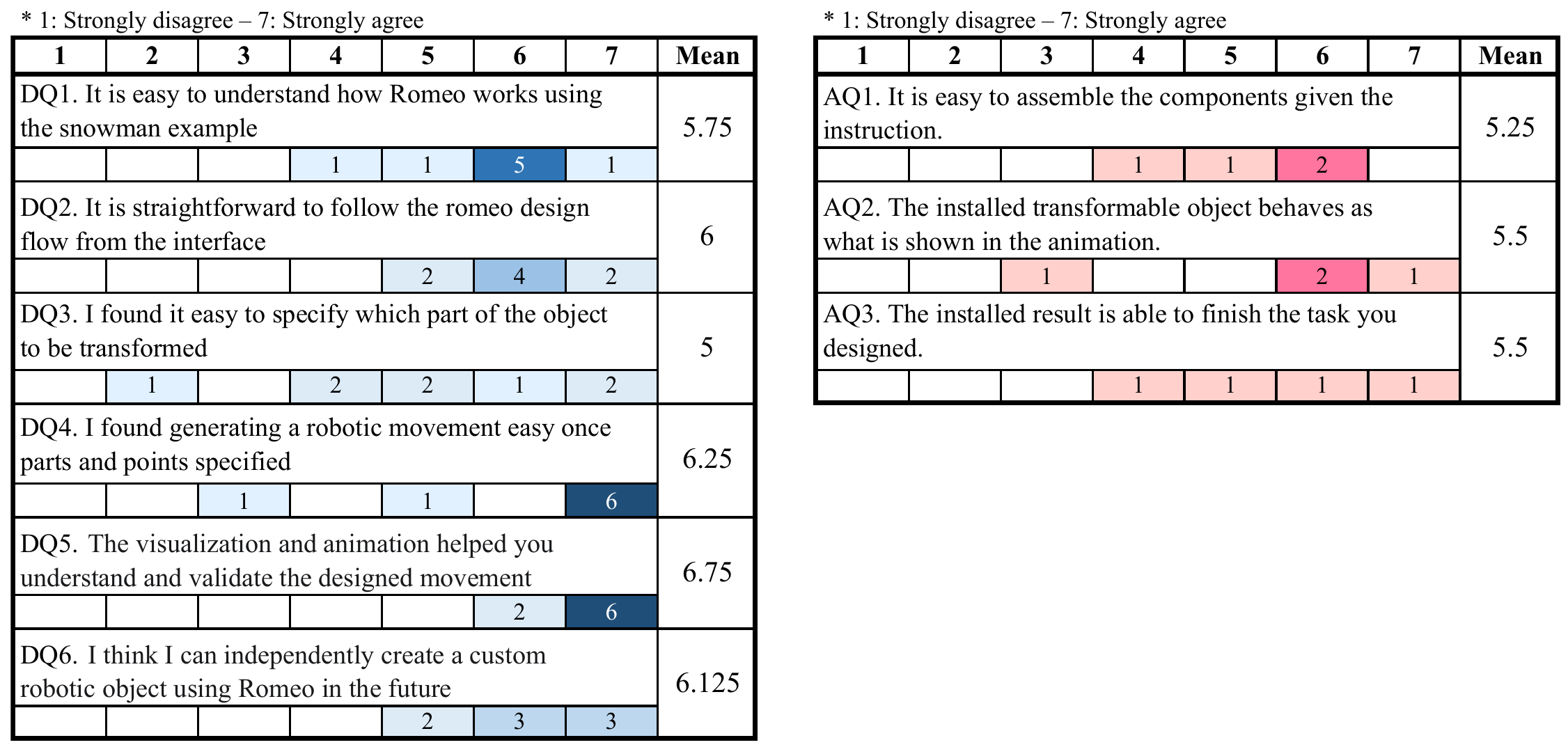}
  \caption{Participants' rating on the Romeo assessment. DQ refers to questions asked in the design session with 8 participants and AQ refers to questions asked in the assembly session with 4 participants
  }
  \label{fig:study-quant}
\end{figure}

\clearpage
\newpage

\begin{figure}[h]
    \centering
    \includegraphics[width=\linewidth]{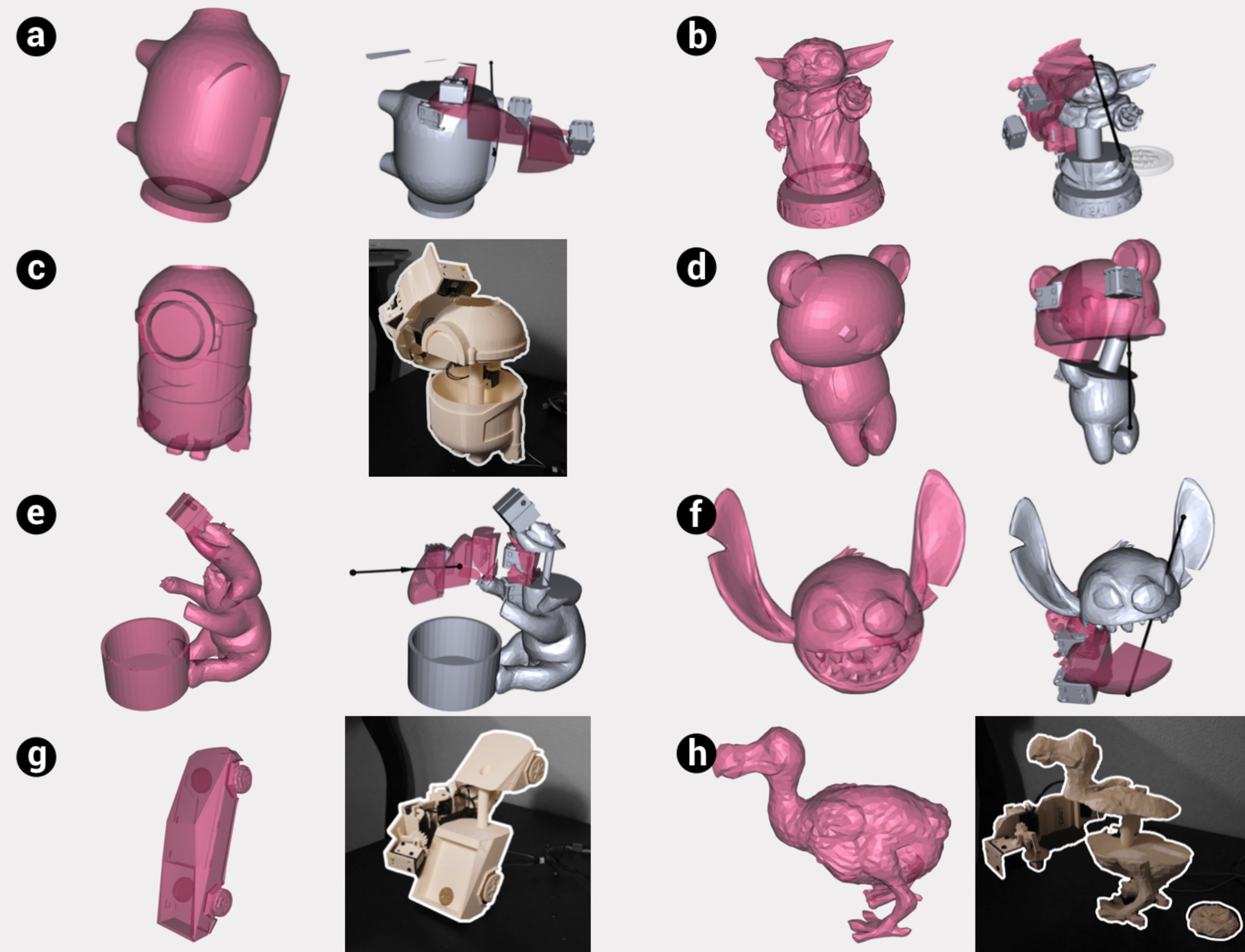}
    \caption{Participants' design of coin-stealing piggy bank of their chosen 3D objects: (a) piggy bank, (b) baby yoda, (c) minion, (d) rilakkuma, (e) polar bears, (f) stitch, (g) Tesla cybertruck and (h) dodo bird. (Original 3D models are designed by Thingiverse users: layerone, MarVinMiniatures, sota919, Anthonylu, MakerBot, Erinfezell, wov, stargatedalek}
    \label{fig:study}
\end{figure}

\end{document}